\documentclass[%
 amsmath,amssymb,
 aps,
pra,
]{revtex4-1}

\usepackage{graphicx}
\usepackage{dcolumn}
\usepackage{bm}

\begin{document}


\title{Stochastic Modelings of Social Phenomena: Pedestrian Counter Flow and Tournaments}

\author{Kichi Tokuyama, Ryuta Maemura, Kengo Yokouchi, Toru Ohira}
  \email{ohria@math.nagoya-u.ac.jp}
\affiliation{%
Graduate School of Mathematics, Nagoya University, Nagoya, Japan
}%


\date{\today}

\begin{abstract}
We present here two examples of stochastic modelings of social phenomena. The first topic is pedestrian counter flow. Two groups of model pedestrians move in opposite directions and create congestions. It will be shown that this congestion becomes worst where individuals are given certain stochastic freedom to avoid another in front compared to the case that they are bound to more strict rules. The second example model tournaments. We present here a rather unexpected feature of tournaments that the probability to reach the top position is higher than that of finishing up at lower positions for not only the number one ranked player, but also for a range of top players. This ``inversion characteristics'' are shown to be observed with simple mathematical model tournaments as well as in the real tournaments.
\end{abstract}

\maketitle


\section{Introduction}
Some of the social systems, such as economic systems, are increasingly becoming of interest of physicists (e.g. \cite{stanley}). These systems are composed of individuals which have non-trivial interactions among themselves. In the modeling, however, these interactions are often approximated much simpler in analogy with physical systems. Also, the detailed complexities are typically treated as stochastic elements in the model.  
Often, collective behaviors of these systems show to rather counter intuitive or complex characteristics. Here, we present our investigation of such examples. The first system is a stochastic modeling of pedestrian counter flows. The second topic deals with a simple modeling of tournaments. Both are based on quite simple rules. However, they lead to interesting phenomena.

\section{Pedestrian Counter Flow}

Among the pedestrian models, which have gained much attention recently, we will study
a counter flow model\cite{nagatani}. Our results, albeit preliminary, suggest rather  counter intuitive relationship between the moves of individuals and the entire flow of gropes\cite{tgf}. In a sense, this example
provides indications of stochastic resonance\cite{bulsara,gam} of flow: With tuned ``noise'' in the motions of individuals, the pedestrian groups show the case of total grid lock, which gives no flow to the desired directions.

Let us describe our model.
Pedestrians are randomly placed on a two dimensional rectangular lattice of 
size $(W \times L) = (150 \times 400)$. The boundary condition is periodic
on all sides (torus). We consider two sets of $N=9000$ pedestrians. One set 
tries to move to the clockwise and the other to the counter clockwise on the lattice of the torus. 
We set $(S_r, S_l)$ as parameters for them to step sideways. At each time step, a pedestrian is chosen to
move one step to one of its four neighboring site by the following rules.
\begin{itemize}
\item
If no one is in front of you in the direction you are heading, he moves one step forward.
\item
 Otherwise, he looks at both his right and left sides. 
\item
If only right (left) side is open and 
$S_r >0$ ($S_l>0$), he moves to the open site.
\item
 If both sides are open, he moves to the right and left site
with the probabilities $P_r = {S_r \over {S_r + S_l}}$ and
$P_l = = {S_l \over {S_r + S_l}}$
\item
If neither site is open, he does not move.
\end{itemize}

With this setting, we performed computer simulations. 
We obtained the following results.

\begin{figure}[h]
\begin{center}
\includegraphics[width=0.7\columnwidth]{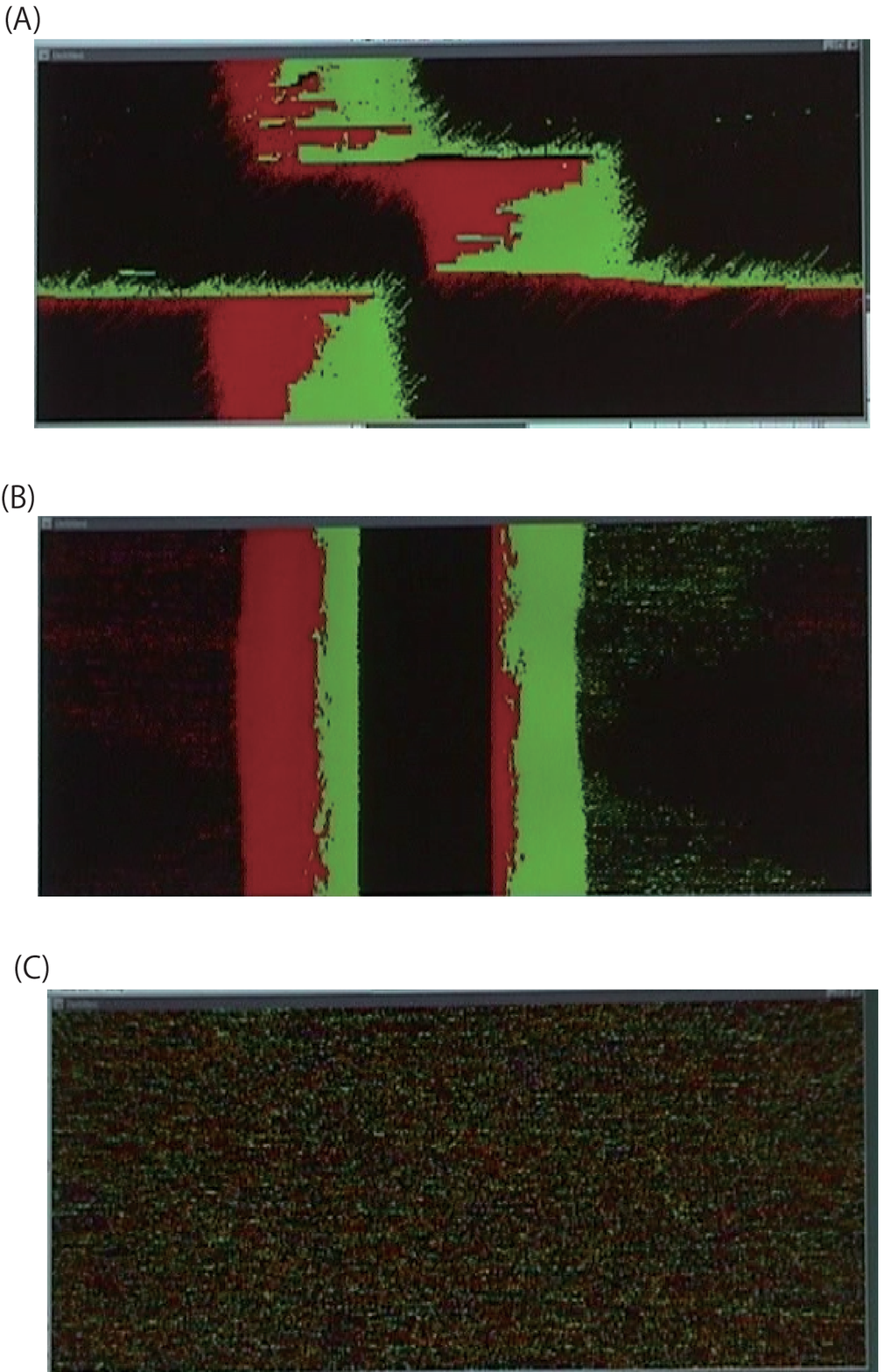}
\end{center}
\caption{Representative results of simulation of pedestrian counter flow. (A) low flow, (B) complete grid lock with no flow, and (C) steady flow.}
\label{pc}
\end{figure}

\begin{itemize}
\item
$(S_r, S_l) = (1.0, 0.0)$.
This is the situation where only forward or to the right moves are allowed.
We typically observe low flow, but not a grid lock, of pedestrians.
(Fig \ref{pc}(A))

\item
$(S_r, S_l) = (1.0, 0.3)$.
The left moves are now allowed. 
However, we typically observe a grid lock situation with no flow.
(Fig \ref{pc}(B))

\item
$(S_r, S_l) = (1.0,0.8)$.
The more left moves are taken. Now, we do not see any
congestion and they are in free flow.(Fig \ref{pc}(C))
\end{itemize}

The point to note is that the flow of the group as a whole is most hindered (as in (B)) not in the case of most strict step rules for individual pedestrians. (Individually, the most strict case is in (A), where one can take a step only to the right.) Mediocre level of freedom for individuals led to the total grid lock as a whole. One should also note that these results depends on density of pedestrians on the torus. Though theoretical understanding of this phenomena is yet to be explored, it can be considered one example of stochastic resonance in collective motions.
\clearpage

\section{Winning and Ranking in Tournaments} 

Tournaments are commonly used in sports and other games. In some sports, such as tennis, there are rankings of players entering into tournaments. It is one of interests of spectators how rankings of players and tournament results compare. Even though there believed to be certain correlations between rankings and winning orders in tournaments, no clear picture has been drawn. By formulating this problem into a mathematical framework, we have found a rather peculiar and counter-intuitive general characteristics: the probability of winning a tournament is highest, compared to that of placed at lower positions, not only for the top ranked player, but also for other high ranked players. There is an indication that this observation is true from the results of real tournaments\cite{tokuyama}.

Let us start by explaining our simple mathematical model tournament with $N$ players. The shape of tournaments is the usual "binary tree-like" with the winner advancing to the next level.
We give each of our players a set $(r, s)$ of a ``rank" $r$ and ``strength" $s$.
At each game in a tournament, the winning probability of a player is set proportional to his relative strength against his opponent. In concrete, in a match by two players $A$ and $B$ with strength of $s_A$ and, $s_B$ respectively, we give the winning
probability for player $A$ equal to $p(A) = {s_A \over {(s_A + s_B)}}$, and similarly for player $B$. (We assume no draw.)

\begin{figure}[h]
\begin{center}
\includegraphics[width=0.7\columnwidth]{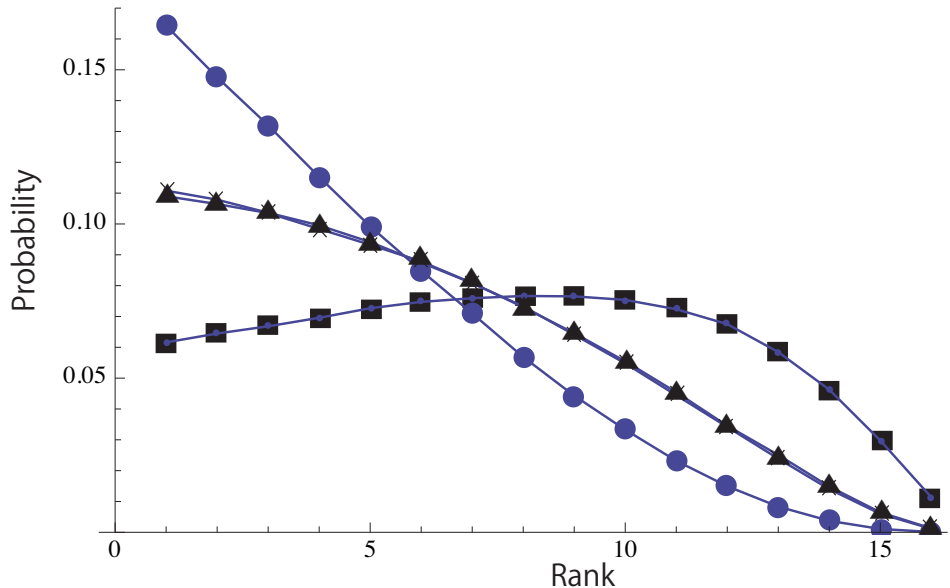}
\end{center}
\caption{The result of the model tournament with $N=16$ players.  The probabilities to finish at the first (dot), second (X), third (triangle), or fourth (square) places are plotted for each player against his rank.}
\label{model1}
\end{figure}

As a first step, we consider the case in which each player has a rank and a strength of $(r, N-r+1), r=1,2,3,\dots, N$. Through combinatorial calculations, we investigated how each player in a tournament finishes. The probabilities for a player to become first, second, third, or fourth places in tournaments against his rank are plotted in Figure \ref{model1}.

The player ranked at the top has the best chance of winning the first place in the tournament, which is as expected. However, the most notable and counter-intuitive point is that for the second to fifth ranked players, their chance of reaching the first place is higher than that for them to be placed at positions according to their ranks. For example, for the third ranked player, the chance that he wins the tournament is better than for him to finish at the second or third places.

Table 1 show that, for different size of tournaments, the range of higher ranked players who have the probability to win the first place higher than that of their becoming of other positions. We see that certain ranges of top players have this ``inversion characteristics'' in the model.
\begin{table}[h]
\begin{center}
\begin{tabular}{|c|c|c|c|c|c|c|c|c|c|c|} \hline
Number of players  & 4 & 8 & 16 & 32 & 64 & 128 & 256 & 512 & 1024  \\ \hline 
Range of inversion & 1-2 & 1-3 & 1-5 & 1-9 & 1-16 & 1-29 & 1-55 & 1-95 & 1-178  \\ \hline
\end{tabular}
\end{center}
\caption{Ranges of top players who have inversion characteristics with varying size of tournaments.}
\end{table}

We can calculate to show that this observed inversion characteristics is not true if the rule is changed in an unrealistic way so that the higher ranked player always wins in a match. In reality, however, details of winning and losing probabilities in matches vary, and it may be that these inversion characteristics are observed commonly in real tournaments. 

\begin{figure}[h]
\begin{center}
\includegraphics[width=0.7\columnwidth]{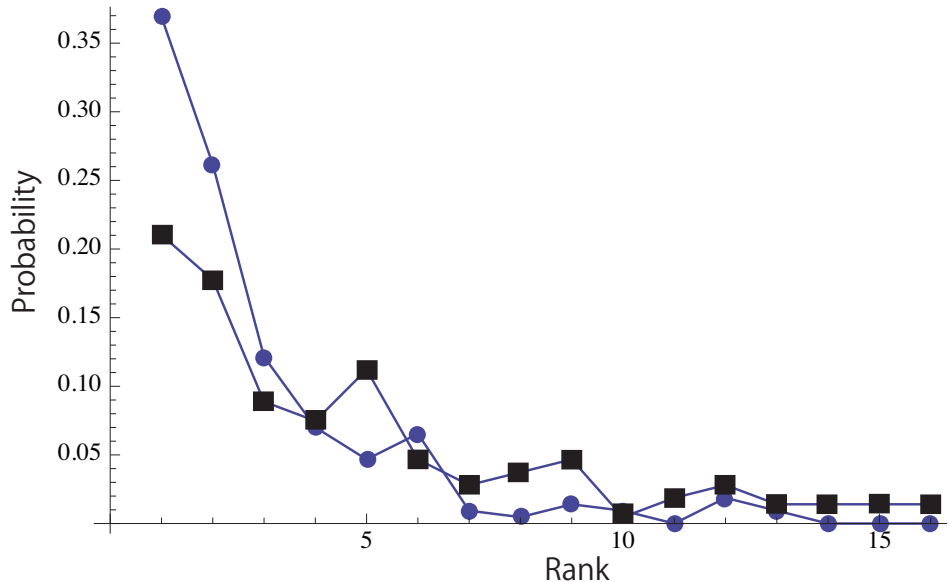}
\end{center}
\caption{A result from real tennis tournaments. Data are taken and compiled from $214$ tournaments with 16-seedings over the period of 1952 to 2001. 
The probabilities to finish at the 
first (dot), or
second (square) places are plotted for each player against his/her rank.(The tournaments did not have matches to decide on the third place.)}
\label{tennis}
\end{figure}

In order to test this hypothesis, we have investigated on real tennis tournaments. Data sets are obtained through the rankings and match results of Association of Tennis Professionals (ATP) \cite{atp} and Women's Tennis Association (WTA) \cite{wta}  The result is shown in Figure \ref{tennis}. Even though statistics are not enough, we can observe the similar inversion characteristics, indicating our hypothesis has certain validity.

There are couple points to note. 
First, we have also considered a hybrid-case where  winning and losing probabilities of national football teams with different strength measured in ``FIFA points''\cite{tokuyama}.   
Statistics are compiled with data of matches 
from Federation of International Football Association (FIFA)  \cite{fifa}. Based on this statistics, we performed a hypothetical tournaments by computer simulations. The inversion characteristics are 
also observed indicating that, regardless of details of winning and losing probabilities in matches, these inversion characteristics are observed commonly in various tournaments
Secondly, in the real tournaments, including the ones shown in Figure \ref{tennis}, we have stronger players placed in certain positions, i.e., seeding. Seedings make the higher ranked players more advantageous in tournaments. Detailed mathematical investigation of such effects is left for further research. We also, note that the probabilities to be second or third places are close together in our simple mathematical model and simulations. This is related to the fact both are the results of losing one game with the same number of matches in a tournament. As many tournaments do not have the third place match, we have not yet investigated whether this can also be seen in the real tournaments.

\section{Discussions}
We have presented simple models of pedestrian counter flows and tournaments. Though simple, they have exhibited rather counter intuitive characteristics. More theoretical investigations are needed to uncover the mechanism of these behaviors.


\end{document}